\documentclass[aps,prl,onecolumn,longbibliography,groupedaddress]{revtex4-1}

\usepackage{url,hyperref,lineno,microtype,subcaption}
\usepackage{graphics}
\usepackage{graphicx}
\usepackage{algorithm}
\usepackage{algcompatible}
\usepackage{float}
\begin{document}

\title{The Forest Fire Model: The subtleties of criticality and scale invariance }
	
\author{Lorenzo \surname{Palmieri}}
\email{l.palmieri16@imperial.ac.uk}
\affiliation{Centre for Complexity Science and Department of Mathematics, Imperial College London, South Kensington Campus, SW7 2AZ, UK}

\author{Henrik Jeldtoft \surname{Jensen}}
\email{h.jensen@imperial.ac.uk}
\affiliation{Centre for Complexity Science and Department of Mathematics, Imperial College London, South Kensington Campus, SW7 2AZ, UK}
\affiliation{Institute of Innovative Research, Tokyo Institute of Technology, 4259, Nagatsuta-cho, Yokohama 226-8502, Japan}

\begin{abstract}
Amongst the numerous models introduced with SOC, the Forest Fire Model (FFM) is particularly attractive for its close relationship to stochastic spreading, which is central to the study of systems as diverse as epidemics, rumours, or indeed, fires. However, since its introduction, the nature of the model's scale invariance has been controversial, and the lack of scaling observed in many studies diminished its theoretical attractiveness. In this study, we analyse the behaviour of the tree density, the average cluster size and the largest cluster and show that the model could be of high practical relevance for the activation dynamics seen in brain and rain studies. From this perspective, its peculiar scaling properties should be regarded as an asset rather than a limitation.
\end{abstract}

\keywords{Forest Fire Model, Critical density, Residence times, Largest Cluster, Scale invariance, Cluster size distribution, Average cluster size, Data Collapse}

\maketitle
\section{Introduction}
Soon after the seminal paper introducing Self-Organised Criticality (SOC) \cite{BTW1987}, it was suggested that examples of SOC could include models describing the spread of activation in a manner reminiscent of forest fires or infectious diseases. The degree to which these models were examples of scale invariance and criticality instantly became a subject of intense debate, see, e.g. \cite{Drossel_Schwabl1992} and \cite{Pruessner2012}. Despite the controversy and indications that the Drossel-Schwabl Forest Fire Model lacks scale invariance \cite{Drossel_Schwabl1992}, the dynamics of the model is of a type that seems of direct relevance to many real systems such as the brain \cite{Tagliazucchi2012} and precipitation \cite{Peters2006}. So if for no other reason, it is still worthwhile to develop a better understanding of the behaviour generated by this kind of stochastic spreading and relaxation dynamics and to develop ways to probe the dynamics which can be applied to data from real systems. 
\\	\\
As the name implies, the approach of SOC is to focus on criticality in the sense of equilibrium statistical mechanics, and for this reason, one typically looks for scale invariance and dynamics that can tune sharply to a critical point. Consequently, broad crossover behaviour is not seen as properly belonging to the SOC paradigm. However, the relevance of a theoretical scientific framework is, in the end, determined by how useful it is for the description and analysis of real systems. Seen from this perspective, it is of importance to bear in mind that exact fine-tuning to a completely scale-invariant state is not always observed in systems that exhibit SOC-like behaviour. This is demonstrated, e.g. by the studies of the size distribution of rain showers \cite{Peters2006}, and the bursts of brain activity measured during fMRI scans \cite{Tagliazucchi2012}. Indeed, neither study finds sharp critical behaviour but, despite the systems being of totally different microscopic nature, both identify similar indications of critical behaviour in terms of approximate power laws and even features reminiscent of peaked, or perhaps diverging, fluctuations or susceptibilities. Moreover, both studies find that the dynamics pulls the systems into a crossover region of large fluctuations within which one most frequently finds configurations to reside. In other words, in both cases, the distribution of residence times, i.e. the amount of time spent at a certain value of the control parameter, is found to exhibit a broad peak centred about what appears to be a critical-like value of the control parameter. From our experience with equilibrium critical phenomena, one would expect a broad peak for small system sizes only, and that the width of the peak will shrink with increasing system size. Therefore, for systems as big as the atmosphere or with as many components as the brain, broad peaks would not be expected. 
	\\ \\ The experimental observations in \cite{Tagliazucchi2012, Peters2006}  suggest that the dynamics couples the control parameter to the fluctuations in a way that makes the system move around in a critical region, rather than tuning to a critical point. This is similar to suggestions previously put forward, such as \cite{Sornette1992, Zapperi1995}. Intuitively, one may imagine something like the following in the case of precipitation: nucleation of drops can happen at a particular value of the vapour content. The vapour in the atmosphere over the ocean gradually builds up towards that value, and sometimes overshooting may even occur before nucleation is seeded. When precipitation events occur, a significant amount of vapour is removed from the atmosphere, and one observes oscillations between subcritical and supercritical regions. For some reason, the coupling between the driving (vapour formation) and the response (precipitation) produces such large fluctuations that a precise tuning to the critical value of the control parameter is excluded. A parallel situation may take place in the brain as neurones have to go through a refractory period before they are able to fire again after discharge.    
\\ \\ A link between the behaviour observed in \cite{Peters2006, Tagliazucchi2012} and the Forest Fire Model was already suggested in \cite{Palmieri2018}. Here we analyse in detail the residence time distribution (which corresponds to the tree density distribution in the FFM) and its relationship to the control parameters used in standard percolation \cite{CM2005} and \cite{Tagliazucchi2012}, namely the mean cluster size and the size of the largest cluster. We find that there is indeed a critical-like region where the fluctuations peak up and where the system seems to spend most of the time, exactly like in the brain and rain observations. Furthermore, analysing the distribution of the largest cluster we find that it displays scale invariance for our reachable sizes of $\theta$ and that the average cluster size seemingly grows superlinearly with $\theta$, in a way that reminds the superlinear growth of the instantaneous correlation length observed in \cite{Palmieri2020}.

\newpage
	\section{Model description}
\begin{figure}[h!]
		\begin{center}
			\includegraphics[scale=0.7]{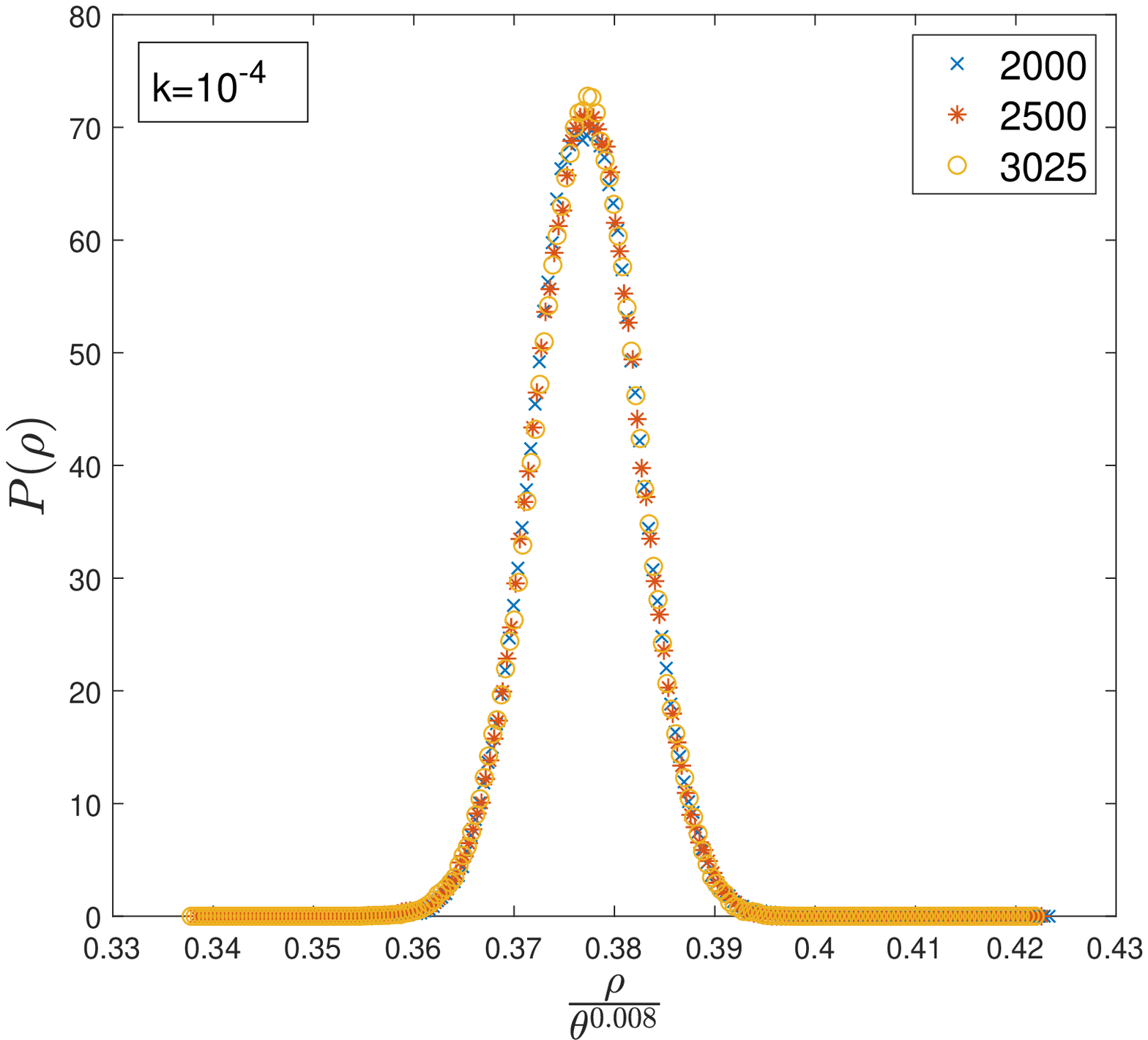}% This is a *.eps file
		\end{center}
		\caption{Distribution of densities $P(\rho)$ computed at different values of $\theta$ and $k=10^{-4}$. Although rescaling the x-axis by $\theta^{0.008}$ seems to make it possible to perform a data collapse for $\theta \in [2000, 3025]$, this doesn't hold for larger values of $\theta$.}\label{k4}
	\end{figure}

\begin{figure}[h!]
		\begin{center}
			\includegraphics[scale=0.7]{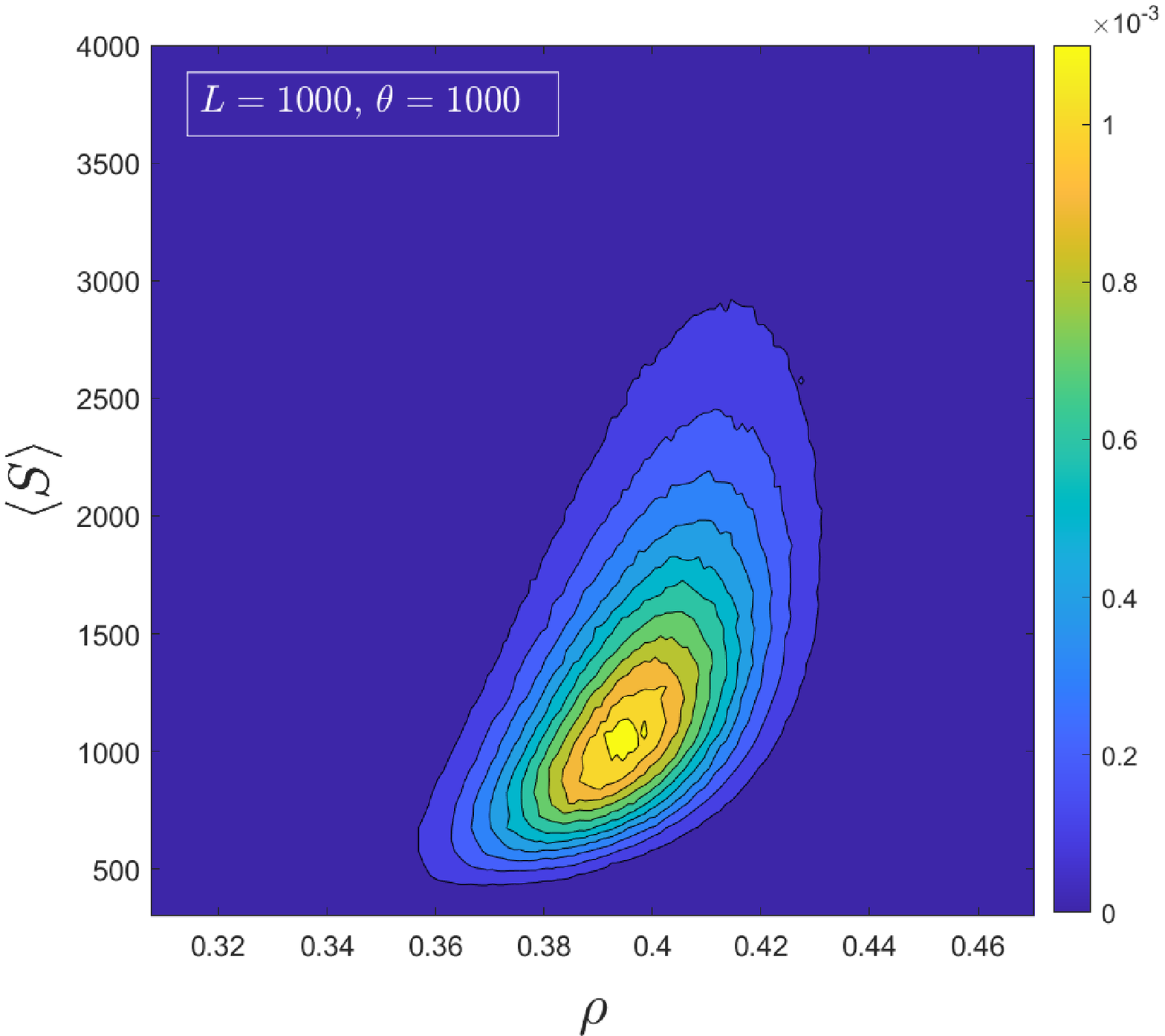}% This is a *.eps file
		\end{center}
		\caption{Heatmap representing the bivariate histogram of the average cluster size and the density.}\label{aves}
	\end{figure}

\begin{figure}[h!]
		\begin{center}
			\includegraphics[scale=0.7]{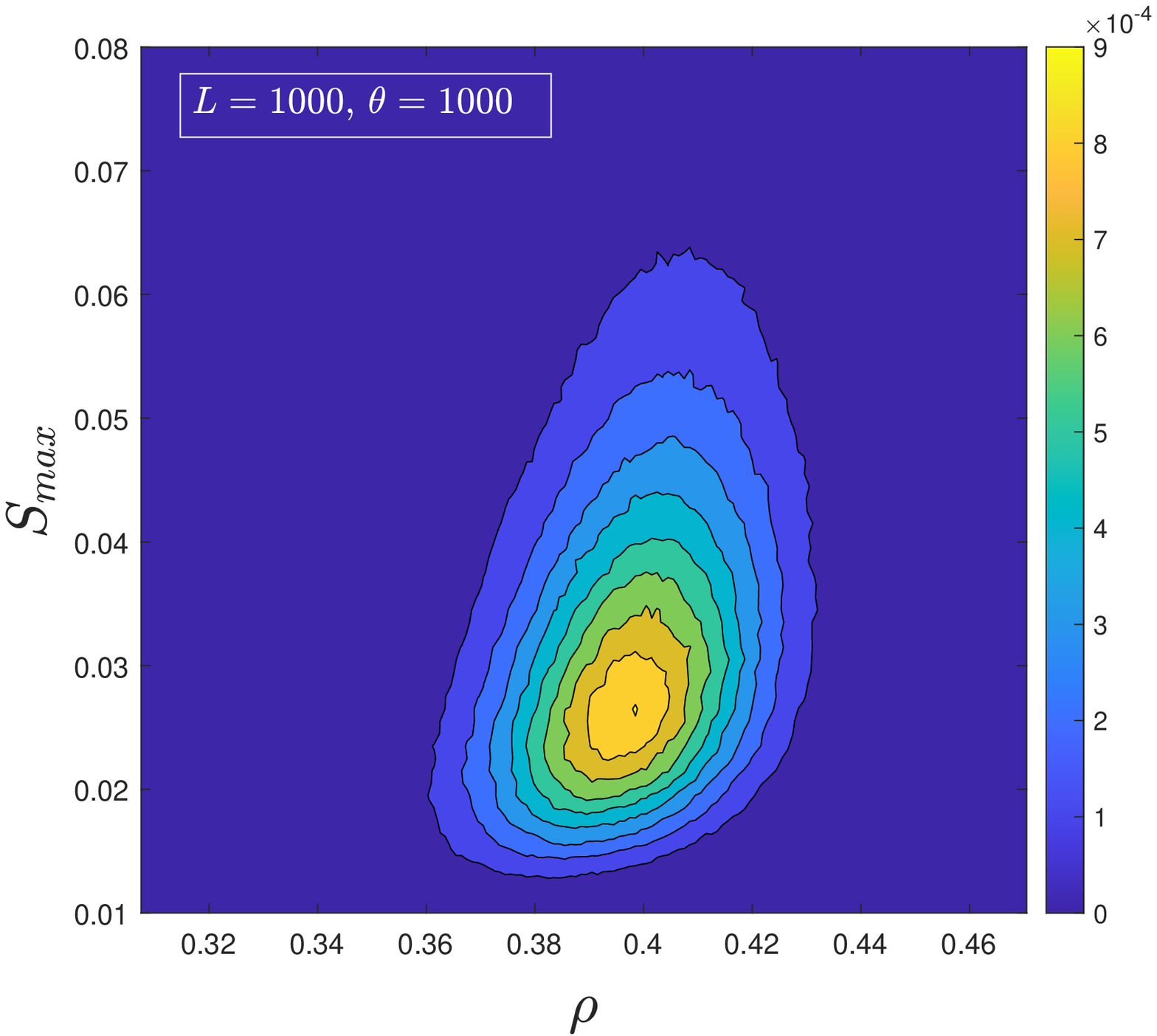}% This is a *.eps file
		\end{center}
		\caption{Heatmap representing the bivariate histogram of the largest cluster size and the density.}\label{maxs}
	\end{figure}

\begin{figure}[h!]
		\begin{center}
			\includegraphics[scale=0.7]{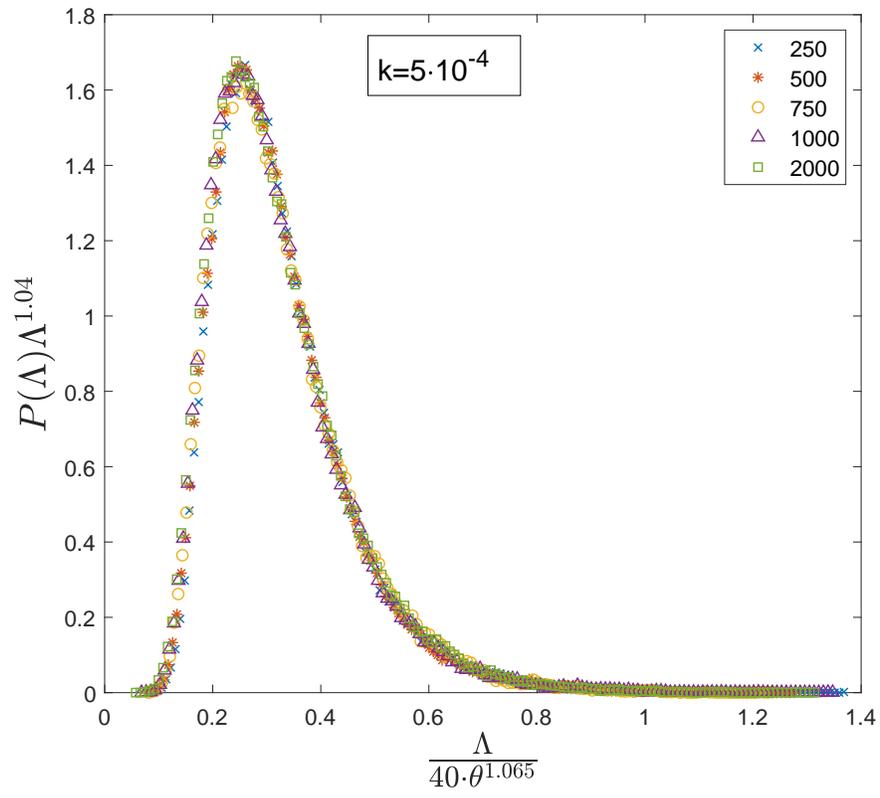}% This is a *.eps file
		\end{center}
		\caption{Data collapse for the distribution of the largest cluster $P(\Lambda)$ at $k=5 \cdot 10^{-4}$. }\label{lambda2}
	\end{figure}

As reported in \cite{Pruessner2012}, the original forest fire model proposed in \cite{Bak1990} was revised in \cite{Drossel_Schwabl1992} and became known in the literature as the Drossel-Schwable Forest Fire Model (DS-FFM), despite the resulting model coinciding with the one introduced a few years before in \cite{Henley1989} as \textit{Self-organised percolation}. In the following, we will analyse the model proposed in \cite{Henley1989, Drossel_Schwabl1992} and refer to it simply as forest fire model (FFM). The FFM consists of a dissipative dynamic that involves the occupation of empty sites (planting of trees) and the removing of clusters of trees (burning of a forest). In the following, we will restrict our simulations to a two-dimensional square grid with periodic boundary conditions and side length $L$. Our implementation follows \cite{Grassberger1993, Clar1994, Schenk2000, Pruessner2002}, and is summarized by the following pseudo-code.
		\begin{algorithm}[H] \label{algo}
		\caption{Forest Fire Model}
		\label{Algo}
		\begin{algorithmic} [0]
			\WHILE{True}
			\FOR{i = 1:$\theta$} \STATE{select randomly a site s} 
			\IF{s is empty} \STATE {s becomes occupied}  \ENDIF 
			\ENDFOR
			\STATE{select randomly a site s}
			\IF{s is occupied}
			\STATE {collect statistics}
			\STATE {burn the cluster connected to s}
			\ENDIF 
			\ENDWHILE
		\end{algorithmic}
	\end{algorithm}
At each time step, we select a site at random; if it is empty, we try to plant $\theta$ trees, if it is occupied, we burn the whole cluster connected to it, considering 4 neighbours for each site. In order to avoid finite-size effects, $\theta$ has to be tuned, taking into account the system size $L^2$. In our simulations we keep fixed the ratio $k=\frac{\theta}{L^2}$ and use periodic boundary conditions as in \cite{Palmieri2020}. The great majority of the simulations are done at $k=10^{-3}$, but we also present results for smaller values of $k$ in Fig. \ref{k4} and in Fig. \ref*{lambda2}. Since the model is known for taking a long time to thermalize, we performed $5*10^6$ burning steps for thermalisation and $10^6$ to collect the statistics as in \cite{Pruessner2002}. The only exception is for the heatmaps in Fig. \ref{aves} and Fig. \ref{maxs}, where $10^7$ datapoints were used to produce the statistics.  
\section{Results}
\subsection{The distribution of densities}
\begin{figure}[h!]
		\begin{center}
			\includegraphics[scale=0.7]{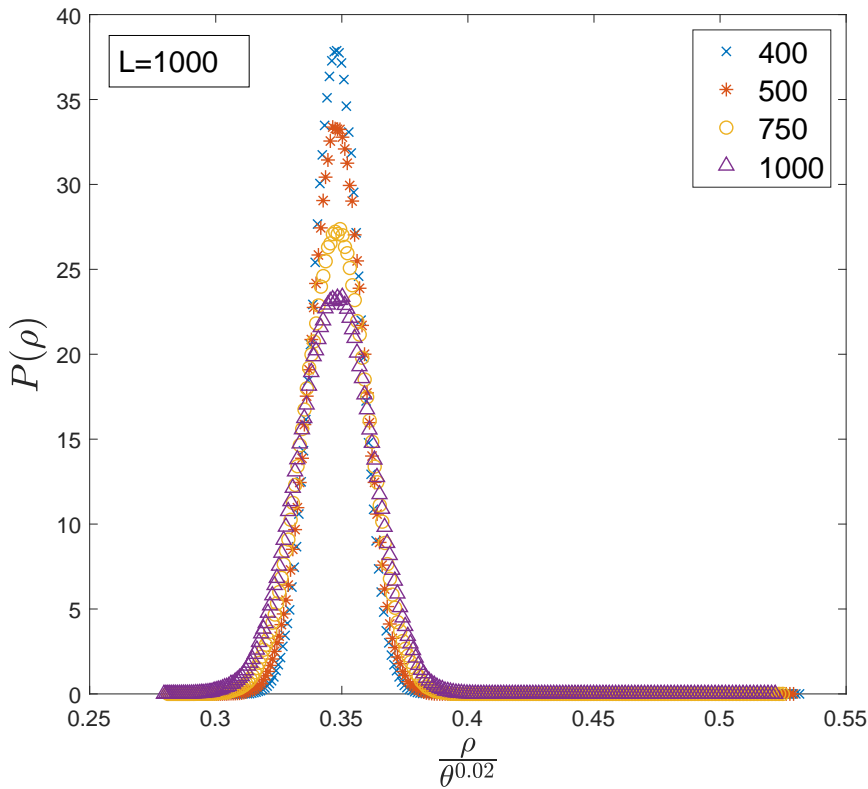}% This is a *.eps file
		\end{center}
		\caption{Distribution of density $P(\rho)$ computed at different values of $\theta$ and fixed $L=1000$. }\label{lfix}
	\end{figure}

\begin{figure}[h!]
		\begin{center}
			\includegraphics[scale=0.7]{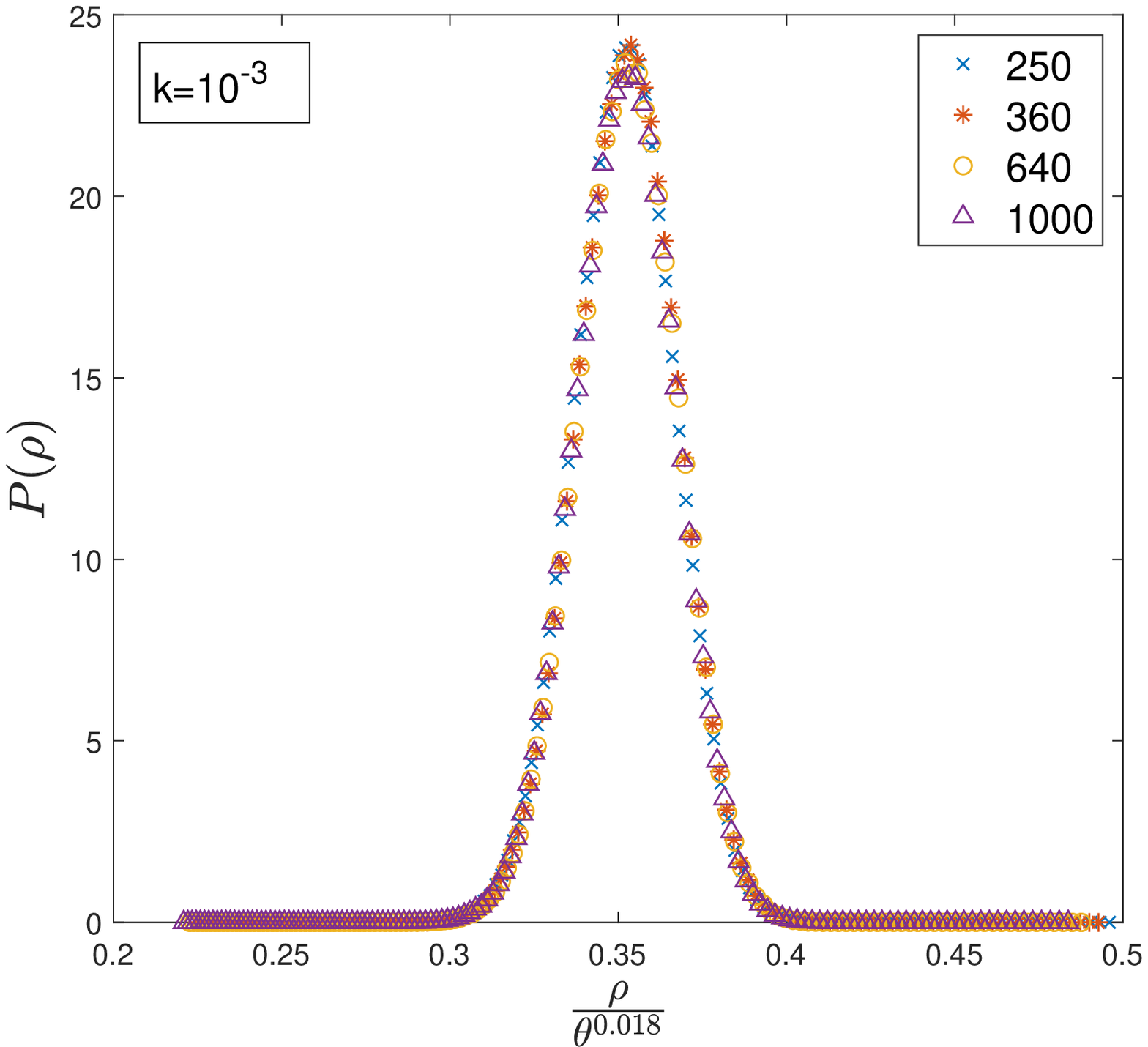}% This is a *.eps file
		\end{center}
		\caption{Distribution of densities $P(\rho)$ computed at different values of $\theta$ and $k=10^{-3}$. Although rescaling the x-axis by $\theta^{0.018}$ seems to make it possible to perform a data collapse for $\theta \in [250, 1000] $, this doesn't hold for larger values of $\theta$.} \label{k3}
	\end{figure}

\begin{figure}[h!]
		\begin{center}
			\includegraphics[scale=0.7]{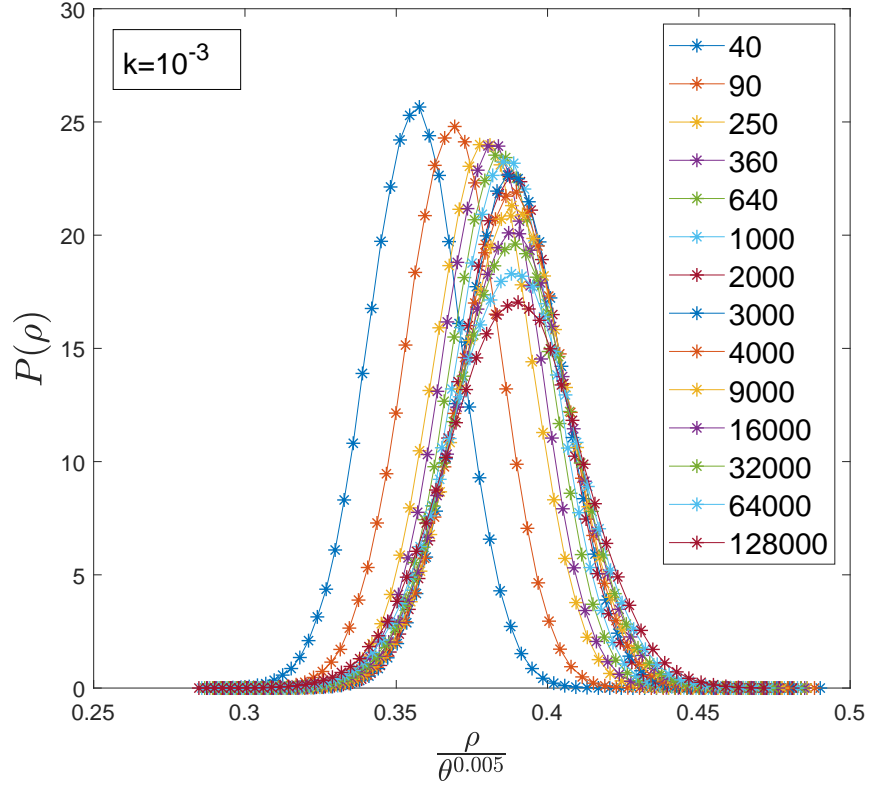}% This is a *.eps file
		\end{center}
		\caption{Distribution of densities $P(\rho)$ computed at different values of $\theta$ and $k=10^{-3}$. The x-axis has been rescaled by $\theta^{0.005}$ to align the position of the peaks at large values of $\theta$.}\label{allk3}
	\end{figure}

\begin{figure}[h!]
		\begin{center}
			\includegraphics[scale=0.7]{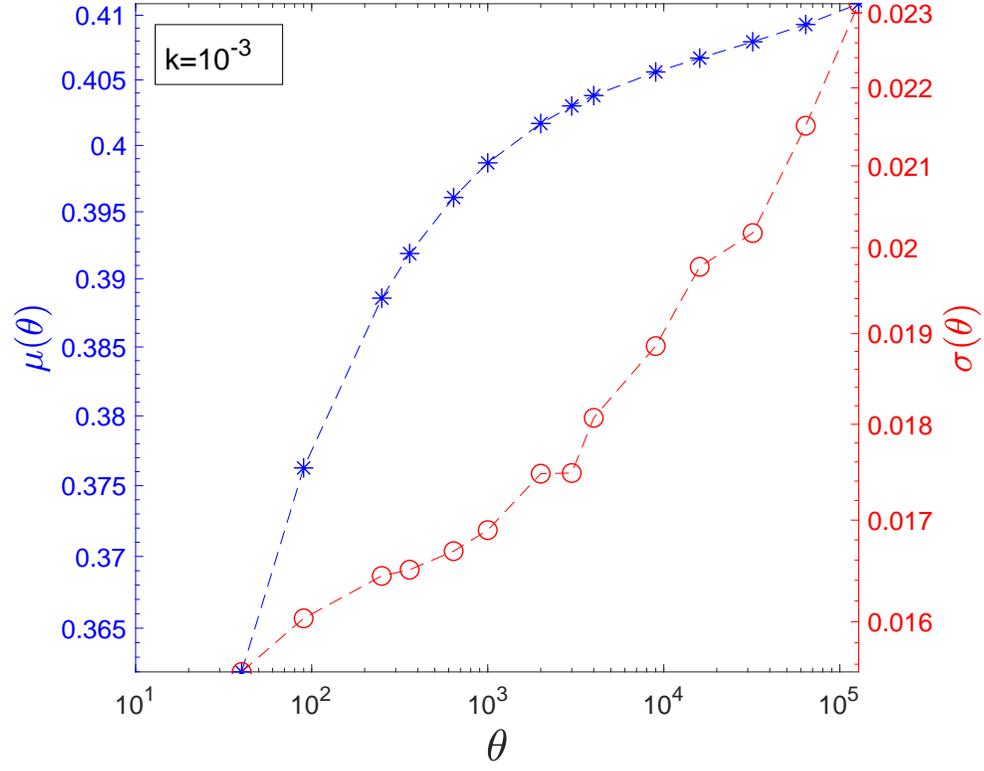}% This is a *.eps file
		\end{center}
		\caption{Mean and standard deviation of $P(\rho)$ as a function of $\theta$.}\label{musi}
	\end{figure}

\begin{figure}[h!]
		\begin{center}
			\includegraphics[scale=0.7]{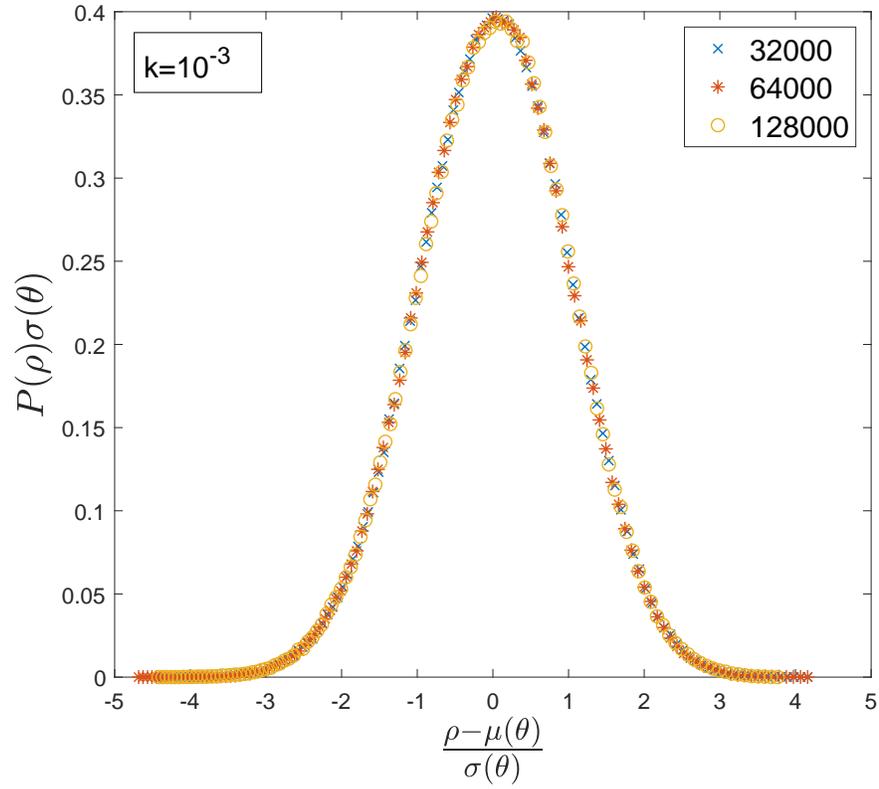}% This is a *.eps file
		\end{center}
		\caption{Data collapse for the distribution of densities $P(\rho)$ and $k=10^{-3}$ obtained by inputting the estimated values of $\mu$ and $\rho$ into a Gaussian distribution. The x-axis and the y-axis have been rescaled in order to make all the curves collapse on a standard normal distribution.}\label{gauss}
	\end{figure}

To compute the distribution of the densities $P(\rho)$, we assign one spatially averaged density $\rho$ to each generated configuration and sample the distribution over the ensemble. In this way, we obtain an object that is equivalent to the distribution of residence times in \cite{Peters2006,Tagliazucchi2012}. In the early studies of the FFM, the average density of trees $\langle \rho \rangle$ was assumed to behave as 
	\begin{equation} \label{avedens}
	\langle \rho \rangle = \rho_{\infty} -a \theta^{-b}
	\end{equation}
The value of the power was estimated as $b=0.47$ in \cite{Honecker1997,Pastor2000} and $b=0.5$ in \cite{Grassberger1993}. However, it was noted in \cite{Grassberger2002} that for $\theta > 10^4$ the average density starts to deviate dramatically from the estimates made at lower values, ending up to be more than 100 standard deviations higher than the expected value for $\theta \sim 10^5$, and seemingly growing as a pure power law for $\theta > 10^4$. Assuming that the power law behaviour holds asymptotically in $\theta$, it was estimated in \cite{Grassberger1993} that the tree density would reach the percolation value ($ \rho_p\simeq 0.5927 \ldots$) for $\theta \sim  10^{40}$.
\\ \\ To avoid finite-size effects, in \cite{Grassberger2002} the values of $\theta$ and $L$ were chosen as follows:  for every value of $\theta$, several simulations were carried out at different system sizes, and it was verified that the distribution of the rms radius,  cluster sizes,  and burning time were independent of $L$. The distribution of densities, on the other hand, depends strongly on the system size. For a fixed value of $\theta$, increasing the system size would make the variance of the distribution decrease as $L^{-2}$ in the absence of finite size effects \cite{Pruessner2002}. Conversely, fixing $L$ and increasing $\theta$ would lead to an increase in the standard deviation and $\langle \rho \rangle$, as can be seen in Fig. \ref{lfix}. In the following, we will focus our attention on the behaviour of $P(\rho)$ for increasing system sizes and fixed values of $k$. 
\\ \\ In Fig. \ref{k3} and Fig. \ref{k4} we plot the distribution of densities rescaling the x-axis by $\theta^\nu$, and tune $\nu$ to align the peaks of the distributions computed at different values of $\theta$. Interestingly, in both cases the curves seem to collapse on the same shape for quite a wide range of $\theta$, suggesting that the distribution has reached a stationary state and that increasing $\theta$ would imply only a shift of $\langle \rho \rangle$. However, increasing $\theta$ even further clearly shows that the asymptotic behaviour suggested by the data collapse is only transient, as the curves start to deviate considerably from the shape observed at lower values of $\theta$, as can be seen in Fig. \ref{allk3}. Rescaling $\rho$ by $\theta ^ {0.005}$ one can align the position of the peaks in Fig. \ref{allk3} for $\theta > 10^4$, but it is clear that, in order to perform a data collapse, the y-axis must be rescaled as well, as the height of the peaks decreases with $\theta$.  
\\ \\ For a fixed $\theta$, we found that $P(\rho)$ is well fitted by a Gaussian distribution. In Fig. \ref{musi}, we plot the mean and standard deviations of $P(\rho)$  for $k=10^{-3}$ and find that the average density seemingly increases as a power-law for $\theta \gtrsim 10^4$, in agreement with the numerics presented in \cite{Grassberger2002}. Although the average $\mu(\theta)$ seems to enter an asymptotic regime after $\theta \gtrsim 10^4$, it is less convincing whether the standard deviation $\sigma(\theta)$ has reached its asymptotic form. 
\\ \\ If we assume a Gaussian behaviour for $P(\rho)$  and that the mean and the standard deviation behave as 
	\begin{equation}
	\mu = a\theta^{\nu_{\mu}} \; {\rm and}\; \sigma = b\theta^{\nu_{\sigma}}.
	\label{mu_and_sigma}
	\end{equation}
then plotting  $P(\rho)  \sigma(\theta)$ versus $\frac{\rho - \mu}{\sigma}$ should produce a standard normal distribution for  systems at $\theta \gtrsim 10^4$. To estimate the asymptotic behaviour, we used the last six data points for $\mu(\theta)$ and the last three data points for $\sigma(\theta)$ in Fig. \ref{musi}, finding 
	\[ a = 0.388 \pm 0.001\;{\rm and}\;  \nu_\mu = 0.0050 \pm 0.0002\]
	and
	\[b = 0.007\pm 0.004\; {\rm and}\;  \nu_{\sigma} = 0.099\pm 0.047\]
	using $95 \%$ confidence bounds. 
\\ \\ Using these estimates,  we find that a data collapse seems to hold well for very large values of $\theta$, as can be seen in Fig. \ref{gauss}. However, such extrapolations have to be taken with great care, firstly because of the small data available for the fit and the not so convincing behaviour of $\sigma(\theta)$, and secondly because the average density can never exceed $1$, meaning that the apparent power-law growth has to stop at some point. 
\\ \\ Since $\nu_{\sigma}>\nu_{\mu}$, the fluctuations grow at a faster rate than the average, as one would expect for a system close to criticality. From the ratio $\frac{\sigma(\theta)}{\mu(\theta)} \sim 0.180 \cdot \theta^{0.094}$, one can obtain a crude estimate for when the fluctuations would become of the same order of magnitude as the average. This would happen at $\theta \sim 10^{17}$ and $\langle \rho \rangle \simeq 0.425$. However, this argument can hardly hold, as it would imply that within two standard deviations, we would have values of the densities that exceed $\rho=1$, which is of course, impossible. Therefore, we can conclude that for $k=10^{-3}$ and within the Gaussian approximation, we can expect the standard deviation to be smaller than the average, despite growing at a faster rate. 
\\ \\ As a reference, we computed a rough estimate of the value $\theta^*$ that would correspond to an average density equal to the percolation threshold $\langle \rho (\theta^*) \rangle = \rho_p$. This gives $\theta^* \sim 10^{37}$, which is consistent with the estimate made in \cite{Grassberger2002} once the confidence bounds on $b$ and $\nu_{\mu}$ are taken into account. However, there is no reason why the asymptotic density of the FFM should be the same as the percolation, as even for very large systems the clusters would still be correlated and therefore intrinsically different from a percolation process. 
\\ \\ Like previous studies of the FFM, we are unable to settle the true asymptotic scaling behaviour of the model, which is still unreachable with today's computers. However, given that effective scaling behaviour in terms of high-quality finite-size collapses can be obtained for very large system sizes and significant ranges of $\theta$, we do think that a detailed phenomenological understanding of the model can be useful as a reference point for the discussion of real systems such as precipitation and brain dynamics. Given this, we present in the next section a careful numerical study of the effective onset region similar to the regions where the precipitation study \cite{Peters2006} and the brain study \cite{Tagliazucchi2012} found a broad peak in the residence times.
\\
\subsection{The most frequently visited region}
Even though the studies of precipitation \cite{Peters2006} and brain activity \cite{Tagliazucchi2012} related their findings to critical behaviour, both observed a remarkable broad onset of the order parameter, which is certainly not the behaviour seen as one approaches the critical point of a second-order phase transition of an infinite system. The absence of a sharp onset of the order parameter is remarkable because, for system sizes like the earth atmosphere or the human brain, one would not expect any significant finite-size effect if the usual phenomenology of equilibrium phase transitions were any guidance. 
\\ \\ While the activation dynamics characteristic of precipitation and brain phenomena are not at all similar to thermal equilibrium dynamics, both systems are at least at a schematic qualitative level similar to the dissipative dynamics of the FFM, with its cycles of loading and discharging. Here we want to investigate further the relationship between the distribution of residence times and the control parameter suggested in \cite{Palmieri2018} and to determine to what extent the FFM exhibits an onset region similar to those observed in rain and brain. If that is the case, we may perhaps take this as indicative of a kind of "universality" different from the stringent universality definition we know from equilibrium systems and more of a pragmatic nature. Needless to say, could such a universality be established, it may be a great help in attempts to classify the behaviour of activation dynamics in complex systems. Furthermore, it could be taken as indicative of a level of emergent behaviour that is independent of the microscopic details, since rain and brain clearly do operate on totally different substrates. 
\\ \\ We are of course just repeating the original hope of SOC research and suggesting that systems of entirely different nature may indeed exhibit similar emergent collective behaviour, even if the dynamics does not operate in a critical state, but rather is found to inhabit a broad region of approximately scale-free nature. Crucial for our suggestion is the observation that the true asymptotic behaviour of the FFM model appears to happen for such enormous systems sizes that they hardly are of relevance to real macroscopic systems. In contrast, the quasi-scale free behaviour observed in the FFM model for intermediate system sizes \cite{Pruessner2002, Grassberger2002} may very well be helpful for the understanding of observations such as those presented in \cite{Peters2006,Tagliazucchi2012}. 
\\ \\ To investigate the presence of a critical-like region in the FFM, we focus on the onset of measures that characterise the clusters of trees and keep track of the frequencies at which the system visits different regions of the parameter space. The order parameter for precipitation \cite{Peters2006} was taken to be the precipitation rate and for the brain \cite{Tagliazucchi2012}, the size of the largest cluster of activated voxels in the fMRI scans - a choice that appears very natural in the light of ordinary percolation analysis. 
\\ \\ In figures Fig. \ref{aves} and Fig. \ref{maxs} we present the contour plot of the bivariate histograms of the average cluster size \cite{CM2005} $\langle S \rangle$ vs the density $\rho$, and of the largest cluster normalised to the number of trees $S_{max}$ vs $\rho$. The colour map represents the probability of observing a certain point in the parameter space, which is the same as the proportion of time spent by the system at that point. To create the heatmaps, we sampled $10^7$ configurations and grouped points with similar probabilities for better visual representation. Therefore, the histograms are not perfectly normalised. The study of both the average cluster size $\langle S\rangle$ and the normalised largest cluster $S_{max}$ is inspired by the resemblance between the clusters of sites occupied by trees and the ordinary geometrical percolation transition. In percolation, either $\langle S\rangle$ and $S_{max}$ are used to construct an order parameter.  
\\ \\ It is clear from the heat-maps that the density around $\rho^*\simeq 0.4$ stands out and that precisely like in the precipitation study \cite{Peters2006}, and even more so in the brain study \cite{Tagliazucchi2012}, $\rho^*$ is indicative of an onset of the order parameter. Furthermore, the region over which the creation of large events happens is very broad, and the system spends a significant fraction of time in the \textit{critical} region. We know from the Gaussianity of $P(\rho)$, see Fig. \ref{gauss}, that this region stays broad for any reachable system size and values of $\theta$. This very much suggests that the FFM model's quasi-tuning to a critical region is a stylistic feature of important relevance. 
\\
\subsection{The distribution of the largest cluster}	
\begin{figure}[h!]
		\begin{center}
			\includegraphics[scale=0.7]{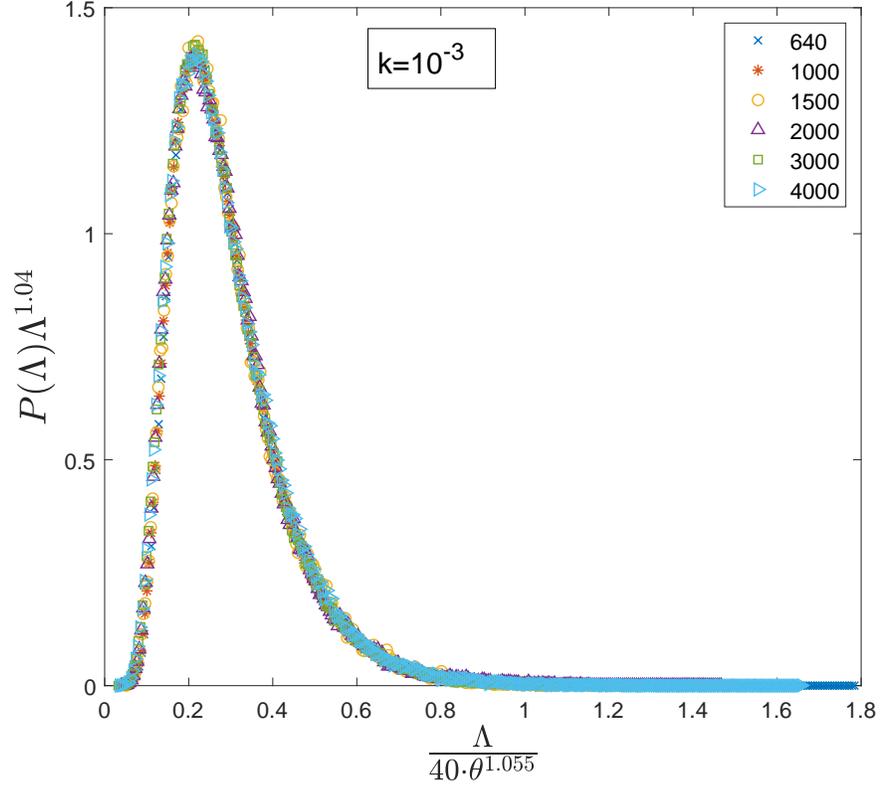}% This is a *.eps file
		\end{center}
		\caption{Data collapse for the distribution of the largest cluster $P(\Lambda)$ at $k=10^{-3}$. }\label{lambda}
	\end{figure}

In the spirit of percolation, we now turn our attention to the non-normalized largest cluster, which we indicate with $\Lambda$ to distinguish it from $S_{max}$. For a fixed system size and $\theta$, we find that the distribution of the largest clusters $P(\Lambda)$ is very well fitted by a Fr{\'e}chet distribution, which corresponds to the class of extreme value statistics with a power law tail. The good agreement of $P(\Lambda)$ with a Fr{\'e}chet distribution implies that the correlations between consecutive configurations generating the clusters are sufficiently weak to be ignored. 
\\ \\Although the distribution of cluster sizes does not obey simple scaling \cite{Pruessner2002, Grassberger2002}, it was found in \cite{Palmieri2020} that the distribution of instantaneous correlation length is scale-invariant for $k=10^{-3}$ and system sizes at least as big as $\theta=9000$. Given that $P(\lambda)$ is fat-tailed, we now turn our attention to its scaling properties and see if simple scaling applies. 
\\ \\ In order to measure $\Lambda$, one has to maintain and keep track of all clusters at all times, which makes the task more computationally intensive than just sampling the density $\rho$. For this reason, we analysed systems sizes that are smaller than those used for the analysis of $P(\rho)$. In our simulations we used $k=10^{-3}$ and values of $\theta$ up to $4000$, but we also checked the scaling for $k = 5 \cdot 10^{-4}$. 
\\ \\ Assuming that simple scaling holds, we can expect $P(\Lambda)$ to follow
\begin{equation} \label{simplescaling}
P(\Lambda) = a G \Bigg( \frac{\Lambda}{\Lambda_{c}(\theta)} \Bigg) \Lambda^{-\tau} 
\end{equation}
for  $\Lambda_0 << \Lambda << \Lambda_c$, where $\Lambda_0$ is a constant lower cutoff and $\Lambda_c$ is an upper cutoff that diverges with $\theta$. In Eq. \ref{simplescaling}, $a$ is a constant metric factor, $\tau$ is the critical exponent, and $G$ is a universal function that decays quickly for $\Lambda>> \Lambda_c$. The form that is usually assumed for the upper cutoff is $\Lambda_c = b \theta ^ \nu$, where $b$ is another constant metric factor and $\nu$ is the spatial dimensionality of the observable $\Lambda$ \cite{Pruessner2012}. 
\\ \\ If the data are consistent with the simple scaling ansatz, then it is possible to perform a data collapse by plotting $P(\Lambda)\Lambda^ \tau$ vs the rescaled variable $\frac{\Lambda}{\Lambda_c}$ for different values of $\theta$. In Fig. \ref{lambda}, we performed a data collapse using $\nu = 0.055$ and $\tau = 1.04$. We also estimated the critical exponents analysing the first two moments of Eq. \ref{simplescaling} and fitting the data with $< \Lambda > = c_1 \theta^{\alpha_1}$ and $< \Lambda^2 > = c_2 \theta^{\alpha_2}$, obtaining
 
	\[ c_1 = 10.8 \pm 0.8\; {\rm and}\;  \alpha_1 = 1.04 \pm 0.01\]
	and
	\[c_2 = 111\pm 37\; {\rm and}\;  \alpha_2 = 2.11\pm 0.04\]
using $95 \%$ confidence bounds.
\\ \\ From the exponents of the first two moments one can easily recover $\nu$ and $\tau$ using Eq. \ref{simplescaling}:
\begin{equation} \label{nutau}
\nu= \alpha_2 - \alpha_1 \quad \tau = 2 - \frac{\alpha_1}{\nu}
\end{equation}
Using our estimates for $\alpha_1$ and $\alpha_2$, we get
\begin{equation} \label{nutau_est}
\nu = 1.07 \pm 0.04\quad \tau = 1.03 \pm 0.04
\end{equation}
using $95\%$ confidence bounds. 
\\ \\ These estimates for $\nu$ and $\tau$ are consistent with the ones obtained via data collapse and that have been used in Fig. \ref{lambda}. Finally, we applied the whole procedure a second time using a smaller value of $k$, namely $k=5 \cdot 10^{-4}$ and $\theta$ up to $2000$. From the data collapse in Fig. \ref{lambda2} we obtained an estimate of $\nu = 0. 065$ and $\tau = 1.04$, while fitting the first two moments we obtained:
	\[ c_1 = 12.1\pm 0.8\; {\rm and}\;  \alpha_1 = 1.05 \pm 0.01\]
	and
	\[c_2 = 152\pm 22\; {\rm and}\;  \alpha_2 = 2.12\pm 0.02\]
which are consistent with the data collapse and with those estimated for $k=10^{-3}$. 
\\ \\ Although we had to restrict our simulations to values of $\theta<10^4$, we observe a very robust scaling for the largest cluster $\Lambda$, which is not observed in the distribution of cluster sizes $P(S)$ over the same range of $\theta$ \cite{Pruessner2002}. Interestingly, we found that $\langle \Lambda \rangle \sim \theta ^ {1.04}$, a super-linear growth which indicates that the correlations in the system increase rapidly with $\theta$. This behaviour is consistent with the analysis of the distribution of instantaneous correlation lengths $P(\xi)$ performed in \cite{Palmieri2020}, where it was found that the average instantaneous correlation length grows as $\langle \xi \rangle \sim \theta^{0.56}$ over the same range of $\theta$. 

	\section{Discussion}
SOC was very much inspired by the successes of the renormalisation group studies of equilibrium critical phenomena of the 1970-ties and its phenomenal understanding of the origin of universality classes. Initially, the discussions of SOC focused on accurately establishing the scaling behaviour and related scaling exponents of the various models thought to represent the SOC phenomenology. The original sandpile model \cite{BTW1987} was disappointingly far from simple scaling, and also the FFM model turned out to behave very differently from the familiar scaling of equilibrium models such as the Ising model or geometrical percolation. Though there is at least one class of SOC models that exhibits clear scaling, namely the class represented by the Manna model \cite{Manna1991}, the beauty of strict scaling and universality classes defined by scaling exponents seems not to really capture the less than ideal critical behaviour frequently observed in real systems, such as our two examples from rain and brain. 
\\ \\ There is no doubt that the studies of the emergent dynamics of real complex systems from biology, geophysics, astrophysics, economics and more \cite{Ball2004,West2017,Krakauer2019} keep identifying behaviour which is qualitatively in the spirit of the hopes and dreams behind SOC, namely lack of one characteristic scale in time and space, large fluctuations and no need for specific external tuning. So if the beauty of strict scaling and exact power laws does not carry over from equilibrium critical phenomena, the question is how we establish a systematic classification of the emergent phenomena observed in completely different systems. The study of the FFM model and its comparison with the behaviour of real systems suggests it is possible to establish a useful phenomenological understanding and classification reaching beyond the usual strict classification of universality classes defined in terms of shared scaling exponents. 
\\ \\ The study presented here confirms earlier investigations of the behaviour of the density $\rho$, and the change in the behaviour of $\langle \rho \rangle$ for very large system sizes. Although we observed that the Gaussian behaviour of $P(\rho)$ holds at least until $\theta = 10^5$, we also showed that it is possible to obtain very good but deceitful data collapses for $P(\rho)$ at different ranges of $\theta$. If, on the one hand, this should serve as a warning for the analysis of the asymptotic scaling behaviour of the FFM and other out of equilibrium systems, on the other hand, it shows that such \textit{effective} scaling could be of guidance in the understanding and analysis of real systems that show similar dynamics. 
\\ \\ Analysing the behaviour of the average cluster size and the largest cluster, we found that the FFM exhibits very similar behaviour to the one observed in experimental studies of rain and brain \cite{Peters2006, Tagliazucchi2012}. In particular, both studies and the FFM display a \textit{critical region} over which both the residence times distribution and the order parameter peak up, meaning that the system spends most of the time in a highly susceptible state.
\\ \\ From the study of the largest cluster $\Lambda$ it emerges that, although the FFM displays broken scaling in the distribution of cluster sizes, the distribution $P(\Lambda)$ appears to be scale-invariant at least for $\theta < 10^4$. Furthermore, $P(\Lambda)$ displays a super-linear growth of the first moment similar to the one observed for the average instantaneous correlation length in \cite{Palmieri2020}. It is clear from the anomalous behaviour of the density $\rho$ that such results should be taken with care, as the model seems to enter a new regime when $\theta > 10^4$. However, the robust scaling observed both in the largest cluster and in the instantaneous correlation length suggests that at least for system sizes below $\theta=10^4$ there is a fast and scale-free growth of the correlations with the \textit{activity} $\theta$. 
\\ \\ Although the FFM does not display the reassuring scaling observed in equilibrium models, its phenomenology appears to summarise elegantly and robustly the emergent dynamics of spreading and recharging seen in such disparate phenomena as rain and brain. Moreover, examples of broken scaling and non-exact powerlaws are ubiquitous in nature and in the scientific literature \cite{Clauset2009} and, for this reason, we believe that the characteristic behaviour of the  FFM should be seen as an asset rather than a limitation of the model. 
	
	\section*{Author Contributions}
 L.P. performed the numerical simulations and produced the figures. Both authors discussed the results of the simulations and jointly wrote the manuscript.
		
\section*{Acknowledgments}
L.P. thankfully acknowledges the High-Performance Computing facilities provided by the Imperial College Research Computing Service, (DOI: 10.14469/hpc/2232). 

	\section*{Funding}
L.P. gratefully acknowledges an EPSRC-Roth scholarship (Award Reference No. 1832407) from EPSRC and the Department of Mathematics at Imperial College London. 
	
\section*{Conflict of Interest Statement}
	%All financial, commercial or other relationships that might be perceived by the academic community as representing a potential conflict of interest must be disclosed. If no such relationship exists, authors will be asked to confirm the following statement: 
	The authors declare that the research was conducted in the absence of any commercial or financial relationships that could be construed as a potential conflict of interest.
	
	\clearpage
	\bibliography{biblio}

\end{document}